\documentclass[preprint,12pt]{elsarticle}  
\usepackage{graphicx} 
\usepackage{amssymb} 
\usepackage{amsmath} 

\journal{Physics Letters B} 

\begin{document} 

\begin{frontmatter} 

\title{Constraints on the threshold $K^-$ nuclear potential from FINUDA 
$^{A}{\rm Z}(K^-_{\rm stop},\pi^-){{^{A}_{\Lambda}}\rm{Z}}$ spectra} 

\author[a]{A.~Ciepl\'{y}} 
\author[b]{E.~Friedman} 
\author[b]{A.~Gal\corref{cor1}} 
\cortext[cor1]{Corresponding author: Avraham Gal, avragal@vms.huji.ac.il} 
\author[a,c]{V.~Krej\v{c}i\v{r}\'{i}k} 
\address[a]{Nuclear Physics Institute, 25068 \v{R}e\v{z}, Czech Republic} 
\address[b]{Racah Institute of Physics, The Hebrew University, 91904 
Jerusalem, Israel} 
\address[c]{Department of Physics, University of Maryland, College Park, 
MD 20742-4111, USA}  

\begin{abstract} 
$1s_{\Lambda}$ hypernuclear formation rates in stopped $K^-$ reactions on 
several $p$-shell targets are derived from hypernuclear formation spectra 
measured recently by the FINUDA Collaboration and are compared with 
calculated $1s_{\Lambda}$ formation rates based on a chirally motivated 
coupled channel model. The calculated rates are about $15\%$ of the derived 
rates, and in contrast with previous calculations depend weakly on the 
depth of the threshold $K^-$ nuclear potential. The $A$ dependence of the 
calculated $1s_{\Lambda}$ rates is in fair agreement with that of the 
derived $1s_{\Lambda}$ rates, showing a slight preference for a deep 
density dependent potential, ${\rm Re}\:V_{K^-}(\rho_0)\sim -$(150--200)~MeV, 
over a shallow potential, ${\rm Re}\:V_{K^-}(\rho_0)\sim -50$~MeV. 
These new features originate from a substantial energy and density dependence 
found for the in-medium subthreshold $K^- n\rightarrow\pi^-\Lambda$ branching 
ratio that enters the hypernuclear formation rate calculations. 
\end{abstract} 


\end{frontmatter} 

\section{Introduction} 
\label{sec:intro} 

How strong is the $K^-$ nuclear interaction? Various scenarios proposed for 
kaon condensation in dense neutron-star matter \cite{KNe86}, and more recently 
for quasibound $K^-$ nuclear clusters \cite{AYa02} and for self-bound strange 
hadronic matter \cite{GFGM} depend on the answer to this 
question which has not been resolved todate. A modern theoretical framework 
for the underlying low-energy $\bar K N$ interaction is provided by the 
leading-order Tomozawa-Weinberg vector term of the chiral effective 
Lagrangian which, in Born approximation, yields a moderately attractive 
$K^-$ nuclear potential $V_{K^-}$: 
\begin{equation} 
\label{eq:chiral} 
V_{K^-}=-\frac{3}{8f_{\pi}^2}~\rho\sim -57~\frac{\rho}{\rho_0},~~~
{\rm (in~MeV)} 
\end{equation} 
where $\rho$ is the nuclear density, $\rho_0=0.17~{\rm fm}^{-3}$, 
and $f_{\pi} \sim 93$ MeV is the pion decay constant. 
This attraction is doubled, roughly, within chirally based coupled-channel 
$\bar K N$--$\pi\Sigma$--$\pi\Lambda$ calculations which provide also for 
a strong absorptivity \cite{WHa08}. Shallower potentials, ${\rm Re}\:V_{K^-}
(\rho_0) \sim -$(40--60) MeV at threshold, are obtained by requiring 
that the in-medium $K^- N$ $t(\rho)$ matrix is derived self-consistently 
with the potential $V_{K^-}=t(\rho)\rho$ it generates \cite{RO00,CFGM01}. 
In contrast, comprehensive global fits to  $K^-$-atom strong-interaction 
shifts and widths yield very deep density dependent $K^-$ nuclear potentials 
at threshold, in the range ${\rm Re}\:V_{K^-}(\rho_0)\sim -$(150--200) MeV 
\cite{FG07}. In this Letter we discuss recent FINUDA measurements that might 
bear on this issue by providing constraints on how deep ${\rm Re}\:V_{K^-}$ 
is at threshold. 

In the preceding Letter \cite{agnello10}, the FINUDA Collaboration at 
DA$\Phi$NE, Frascati, reported on $\Lambda$-hypernuclear excitation spectra 
taken in the 
$K^{-}_{\rm stop}+{^{A}{\rm Z}}\rightarrow \pi^{-}+{^{A}_{\Lambda}{\rm Z}}$ 
reaction on several $p$-shell nuclear targets. Formation rates were given 
per stopped $K^-$ for bound states and for low lying continuum states. In 
$^{16}_{~\Lambda}{\rm O}$ the bound state formation rates agree nicely with 
a previous KEK measurement \cite{tamura94}. The recent FINUDA data allow for 
the first time to consider the $A$ dependence of the formation rates in detail 
within the nuclear $p$ shell where nuclear structure effects may be reliably 
separated out. It is our purpose in this companion Letter to apply one's 
knowledge of the nuclear structure aspect of the problem in order to extract 
the dynamical contents of the measured formation rates, particularly that 
part which concerns the $K^-$ nuclear dynamics at threshold. In doing so 
we transform the partial formation rates reported for well defined and 
spectroscopically sound final $\Lambda$ hypernuclear states into 
$1s_{\Lambda}$ hypernuclear formation rates that allow direct comparison 
with DWIA calculations. 

The expression for the formation rate of hypernuclear final state $f$ 
in capture at rest on target g.s. $i$, apart from kinematical factors, 
is a product of two dynamical factors \cite{CFGM01,GK86,MY88,KCG10}:
(i) the branching ratio for 
$K^- n \rightarrow \pi^- \Lambda$ in $K^{-}$ absorption at rest in 
the nuclear medium, here denoted BR; and (ii) the absolute value 
squared of a DWIA amplitude  given by 
\begin{equation} 
T_{fi}^{\rm DWIA}({\bf q}_{f})=\int \chi^{(-)*}_{{\bf q}_{f}}({\bf r})\:
\rho_{fi}({\bf r})\:\Psi_{nLM}({\bf r})\:d^3r, 
\label{eq:DW} 
\end{equation} 
divided for a proper normalization by the integral $\overline \rho$ of 
the $K^-$ atomic density overlap with the nuclear density $\rho(r)$ 
\begin{equation} 
\overline \rho = \int \rho(r) \mid \Psi_{nLM}({\bf r})\mid^{2}\:d^3r. 
\label{eq:rhobar} 
\end{equation} 
Here $\rho_{fi}$ stands for the nuclear to hypernuclear transition form 
factor, $\chi^{(-)}_{{\bf q}_{f}}$ is an outgoing pion distorted wave 
generated by a pion optical potential fitted to scattering data, and 
$\Psi_{nLM}$ is a $K^{-}$ atomic wavefunction obtained by solving the 
Klein-Gordon equation with a $K^{-}$ nuclear strong interaction potential 
$V_{K^-}$ added to the appropriate Coulomb potential. The integration on 
the r.h.s. of Eq.~(\ref{eq:DW}) is confined by the bound-state form factor 
$\rho_{fi}$ to within the nucleus, where $\Psi_{nLM}$ is primarily determined 
by the strong-interaction $V_{K^-}$, although $\Psi_{nLM}$ is an atomic 
wavefunction that peaks far outside the nucleus. The sensitivity of the DWIA 
amplitude Eq.~(\ref{eq:DW}) to $V_{K^-}$ arises from the interference of 
$\Psi_{nLM}$ with the pion oscillatory distorted wave 
$\chi^{(-)}_{{\bf q}_{f}}$. In particular, once $V_{K^-}$ is sufficiently 
deep to provide a strong-interaction bound state for a given $L$, the atomic 
$\Psi_{nLM}$ also becomes oscillatory within the nucleus which magnifies the 
effects of interference, as verified in past DWIA 
calculations \cite{CFGM01,KCG10}. 

In this Letter we point out another strong sensitivity to the initial-state 
$K^-$ nuclear dynamics arising from the energy and density dependence of the 
$K^- n \rightarrow \pi^- \Lambda$ BR. We show how to incorporate this energy 
and density dependence into the calculation of a properly averaged value 
$\overline{\rm BR}$ which depends on the $K^-$ atomic orbit through $L$ and 
on the mass number $A$ of the target. The resulting calculated $1s_{\Lambda}$ 
formation rates are then compared to those derived from the FINUDA data and 
conclusions are made on the deep vs. shallow $K^-$ nuclear potential issue.

\section{Derivation of $1s_{\Lambda}$ capture rates from FINUDA data} 
\label{sec:finuda} 

The FINUDA spectra show distinct peaks for several $1s_{\Lambda}$ and 
$1p_{\Lambda}$ states in the nuclear $p$ shell. In general, the derivation 
of the $1p_{\Lambda}$ formation rate is ambiguous given that the 
$1p_{\Lambda}$ formation strength is often obscured by a rising $\Lambda$ 
continuum. In $_{\Lambda}^{9}{\rm Be}$ and in $_{~\Lambda}^{13}{\rm C}$ 
it is also mixed with a substantial part of the $1s_{\Lambda}$ formation 
strength owing particularly to high lying $T=1$ parent states in the 
corresponding core nuclei. For this reason, we here deal only with the 
$1s_{\Lambda}$ formation strength, deriving it in each $p$-shell $\Lambda$ 
hypernucleus from unambiguously identified {\it low lying} $1s_{\Lambda}$ 
states. According to Ref.~\cite{DG78}, the corresponding hypernuclear 
formation rates are given by a $1s_{\Lambda}$ formation rate 
$R(1s_{\Lambda})$, which is independent of the particular hypernuclear 
excitation considered, times a structure fraction derived from neutron pick-up 
spectroscopic factors in the target nucleus. This theoretical framework is 
also applicable to forward cross sections of in-flight reactions such as 
$(\pi^+,K^+)$ and $(e,e'K^+)$. In Table~\ref{tab1} we present $1s_{\Lambda}$ 
formation rates derived from the FINUDA $K^-$ capture at rest hypernuclear 
spectra \cite{agnello10,agnello05} for a procedure denoted (a). In each 
spectrum we focus on the strongest low-lying particle-stable hypernuclear 
excitation which is also well described in terms of a $\Lambda$ hyperon weakly 
coupled to a nuclear core parent state. These core parent states are listed in 
the table. The measured formation rates for the corresponding hypernuclear 
excitations from Refs.~\cite{agnello10,agnello05} are then divided by the 
structure fractions listed in the table to obtain values of $R(1s_{\Lambda})$. 
For comparison, we display in the last column the $1s_{\Lambda}$ component of 
forward-angle integrated $(\pi^+,K^+)$ cross sections, also derived using the 
peaks listed in the second and third columns. These $(\pi^+,K^+)$ strengths 
show little $A$ dependence, in contrast to the $K^-$ capture at rest 
$1s_{\Lambda}$ formation rates that decrease by a factor 3.5 in going from 
${^7{\rm Li}}$ to ${^{16}{\rm O}}$. 

\begin{table}[hbt] 
\begin{center} 
\caption{$1s_{\Lambda}$ formation rates $R(1s_{\Lambda})$ per stopped 
$K^-$, derived from the strongest hypernuclear bound state peak for 
each of the listed targets [procedure (a)]. Data are taken from the 
preceding Letter \cite{agnello10}, and for $^{12}_{~\Lambda}\rm{C}$ 
from \cite{agnello05}. The errors are statistical and systematic, 
in this order. The $1s_{\Lambda}$ structure fractions are from 
\cite{MGDD85} and, if unlisted there, from \cite{DG78}. 
Listed in the last column, for comparison, are $1s_{\Lambda}$ 
forward-angle integrated $(\pi^+,K^+)$ cross sections, also derived 
by using procedure (a) from KEK-E336 measurements \cite{hashimoto06}.} 
\begin{tabular}{cccccc} 
\hline 
target & peak & $E^{\ast}_{\rm core}$ & $1s_{\Lambda}$ & 
$R(1s_{\Lambda})\times 10^{3}$ & $\sigma_{1s_{\Lambda}}(\mu b)$ \\  
$^{A}{\rm Z}$ & $J^{\pi}_{\rm core}$ & (MeV) & frac. & per stopped $K^-$ & 
$(\pi^+,K^+)$ \\  \hline 
$^{7}\rm{Li}$ & $3^+$ & 2.19 & 0.311 & $1.48\pm 0.16\pm 0.19$ & 
$1.56\pm 0.10$ \\ 
$^{9}\rm{Be}$ & $2^+$ & 2.94 & 0.242 & $0.87\pm 0.08\pm 0.12$ & 
$1.40\pm 0.05$ \\ 
$^{12}\rm{C}$ & $(3/2)^-$ & g.s. & 0.810 & $1.25\pm 0.14\pm 0.12$ & 
$1.78\pm 0.04$ \\ 
$^{13}\rm{C}$ & $2^+$ & 4.44 & 0.224 & $0.85\pm 0.09\pm 0.13$ & 
$1.87\pm 0.09$ \\ 
$^{16}\rm{O}$ & $(3/2)^-$ & 6.18 & 0.618 & $0.42\pm 0.06\pm 0.06$ & 
$1.47\pm 0.05$ \\ 
\hline 
\end{tabular} 
\label{tab1} 
\end{center} 
\end{table} 

\newpage

In the second procedure, denoted (b) and presented in Table~\ref{tab2}, 
we consider all the particle-stable $1s_{\Lambda}$ states corresponding 
to observed peaks for which the shell model offers reliable identification. 
For three of the five targets listed, this procedure saturates or is close 
to saturating the $1s_{\Lambda}$ formation strength. However, in both 
${^9_{\Lambda}{\rm Be}}$ and ${^{13}_{~\Lambda}{\rm C}}$ the $1s_{\Lambda}$ 
particle stable hypernuclear states represent less than half of the full 
$1s_{\Lambda}$ strength. In the last column of Table~\ref{tab2} we assembled 
$1s_{\Lambda}$ forward-angle integrated $(\pi^+,K^+)$ cross sections, derived 
this time by applying procedure (b). Similarly to Table~\ref{tab1}, the weak 
$A$ dependence of these $1s_{\Lambda}$ $(\pi^+,K^+)$ cross sections is in 
stark contrast to the fast decrease of the $1s_{\Lambda}$ formation rates, 
again by a factor 3.5, going from ${^7{\rm Li}}$ to ${^{16}{\rm O}}$ in 
$K^-$ capture at rest. 
The strong $A$ dependence of the $(K^-_{\rm stop},\pi^-)$ rates with 
respect to the weak $A$ dependence of the $(\pi^+,K^+)$ cross sections 
reflects the sizable difference between the strongly attractive $K^-$ nuclear 
interaction at threshold and the weakly repulsive $K^+$ nuclear interaction. 

\begin{table}[hbt] 
\begin{center} 
\caption{Same as in Table~\ref{tab1} except for using several (rather than 
one) well defined $1s_{\Lambda}$ bound states for each of the listed targets 
[procedure (b)].} 
\begin{tabular}{ccccc} 
\hline 
target & peaks & $1s_{\Lambda}$ & $R(1s_{\Lambda})\times 10^{3}$ & 
$\sigma_{1s_{\Lambda}}(\mu b)$ \\ 
$^{A}{\rm Z}$ & & frac. & per stopped $K^-$ & $(\pi^+,K^+)$ \\  \hline 
$^{7}\rm{Li}$ &1,2,3&0.833& $1.25\pm 0.14\pm 0.17$&$1.29\pm 0.12$ \\ 
$^{9}\rm{Be}$ &1,2&0.435& $0.85\pm 0.09\pm 0.11$ & $1.20\pm 0.05$ \\ 
$^{12}\rm{C}$&1,2,3&0.995&$1.67\pm 0.23\pm 0.23$& $1.92\pm 0.07$ \\ 
$^{13}\rm{C}$&1,2&0.347& $0.84\pm 0.12\pm 0.12$ & $1.93\pm 0.12$ \\ 
$^{16}\rm{O}$&1,2&1.000& $0.36\pm 0.06\pm 0.05$ & $1.32\pm 0.05$ \\ \hline 
\end{tabular} 
\label{tab2} 
\end{center} 
\end{table} 

It is encouraging to see that both sets of $R(1s_{\Lambda})$ values 
in Tables~\ref{tab1} and \ref{tab2} are consistent within statistical 
uncertainties with each other, except marginally for $^{12}\rm{C}$ 
which dates back to a separate FINUDA run \cite{agnello05}. 
Procedure (a) yields a value for $^{12}\rm{C}$ that compares well with 
$R(1s_{\Lambda})[^{12}\rm{C}]=(1.11\pm 0.14)\times 10^{-3}$ per $K^-_{\rm 
stop}$, the latter value corresponding to $E_x \lesssim 7$~MeV in the KEK 
${^{12}_{~\Lambda}{\rm C}}$ spectrum.{\footnote{We thank Dr. Tamura 
for providing details from his Ph.D. thesis \cite{tamura10} on the 
KEK experiment \cite{tamura94}.}} Therefore, in the present study 
we adopt the $R(1s_{\Lambda})$ values listed in Table~\ref{tab1}.

\section{Energy and density dependent $K^- n \rightarrow \pi^- \Lambda$ 
branching ratios} 
\label{sec:BR} 

\begin{figure}[htb] 
\begin{center} 
\includegraphics[width=0.47\textwidth]{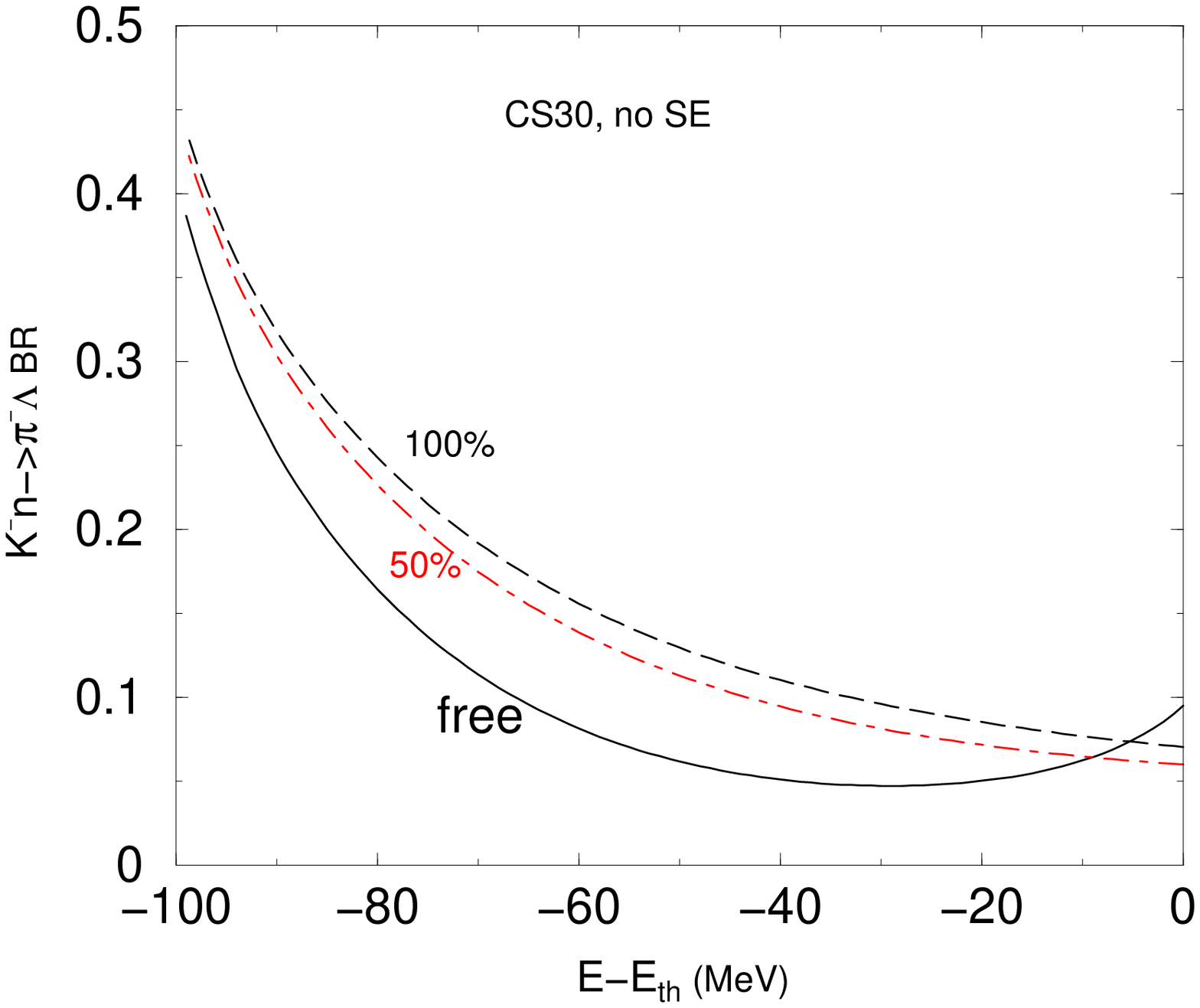} 
\hspace{3mm} 
\includegraphics[width=0.47\textwidth]{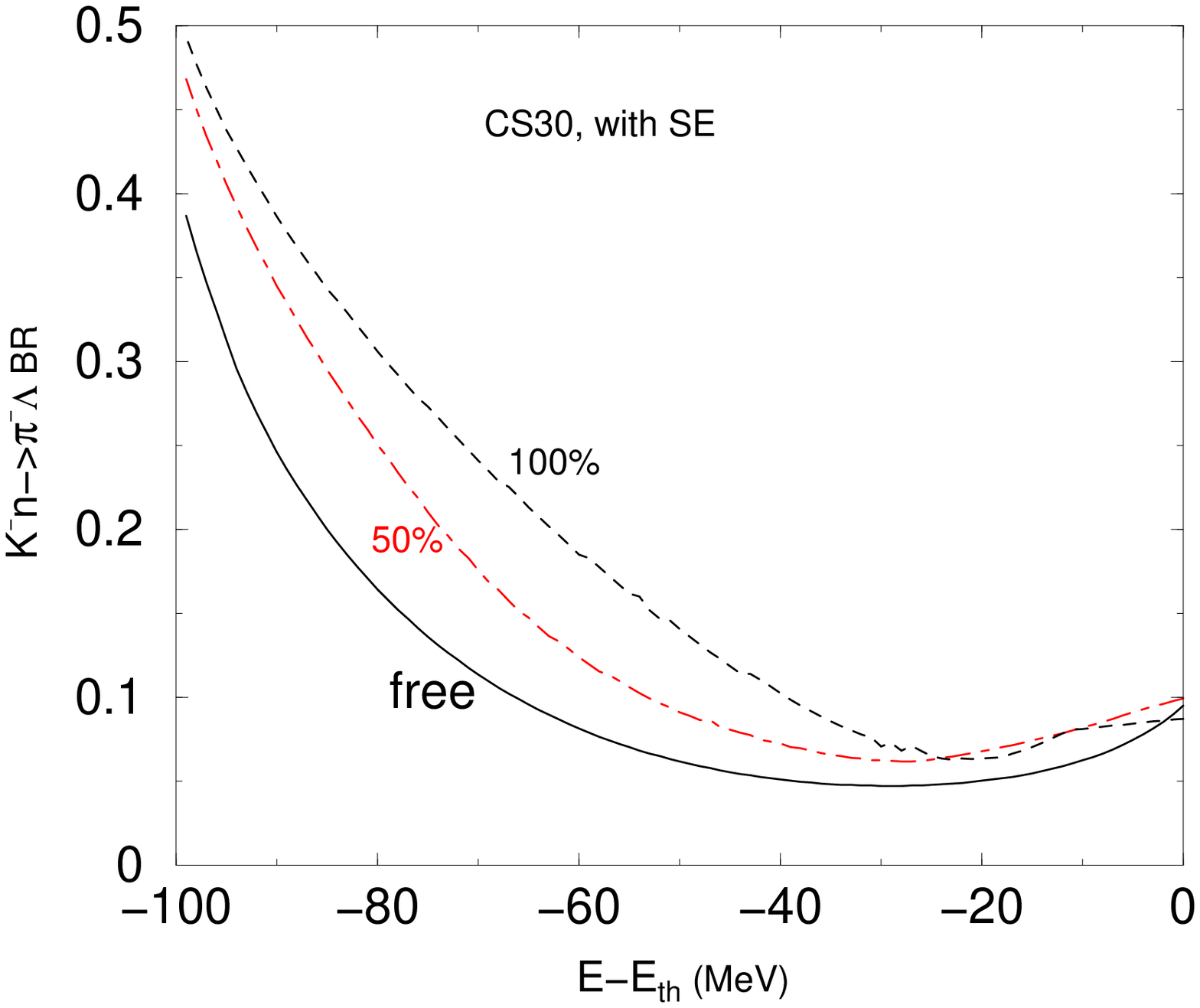} 
\caption{Subthreshold energy dependence of the $K^-n\rightarrow\pi^-\Lambda$ 
branching ratio BR in the CS30 version of the chirally motivated model 
Ref.~\cite{CS10}. The l.h.s. curves for $50,100\%$ nuclear matter density 
demonstrate Pauli blocking effects whereas the r.h.s. curves account 
additionally for self energy effects.} 
\label{fig1} 
\end{center} 
\end{figure} 

Figure \ref{fig1} shows the subthreshold energy dependence of the free-space 
$K^- n \rightarrow \pi^- \Lambda$ BR generated by the CS30 version of the 
chirally motivated coupled channel model of Ref.~\cite{CS10}.{\footnote{The 
parameters of CS30 are constrained by $\sigma_{\pi N}=30$ MeV.}} This $I=1$ 
BR is about $10\%$ at threshold, decreasing to roughly half of its value as 
the $I=0$ $\Lambda(1405)$ subthreshold resonance is traversed, and then 
increases to approximately $40\%$ on approaching the $\pi\Sigma$ threshold 
about 100 MeV below the $\bar K N$ threshold. The figure also shows the 
in-medium BR below threshold at densities $50\%$ and $100\%$ of nuclear 
matter density $\rho_0$, in two versions of medium modifications. In the 
l.h.s. plots the only medium effect is Pauli blocking, which acts in 
intermediate $\bar K N$ states in the coupled channel equations. This is 
known to have the effect of pushing the dynamically generated $\Lambda(1405)$ 
to energies above threshold \cite{koch94,WKW96}, thus weakening the $I=0$ 
interaction and consequently increasing the $I=1$ BR. The energy dependence 
in the subthreshold region is seen to be monotonic. The r.h.s. plots include 
in addition to Pauli blocking also meson and baryon self-energy (SE) terms 
in intermediate state propagators. This pushes back the $\Lambda(1405)$ 
\cite{RO00,lutz98} and in the chirally based model used here \cite{CS10} 
results in stronger energy and density dependencies. The same chiral model was 
used in Ref.~\cite{KCG10} to generate a $K^- n \rightarrow \pi^- \Lambda$ BR 
which, however, was fixed at its threshold value, thus neglecting any possible 
energy dependence. Since the in-medium BRs plotted in Fig.~\ref{fig1} exhibit 
a sizable energy and density dependence, it is essential to consider the 
implied effects in the evaluation of the $1s_{\Lambda}$ formation rates. 

The $K^- n \rightarrow \pi^- \Lambda$ 
BR depends on the initial $K^- n$ invariant energy $\sqrt{s}$, with 
$s=(E_K+E_N)^2-({\vec p}_K+{\vec p}_N)^2$ in obvious notation. In the two-body 
c.m. system ${\vec p}_K+{\vec p}_N = 0$, but in the $K^-$--nucleus c.m. 
system (approximately nuclear lab system) ${\vec p}_K+{\vec p}_N \neq 0$ and 
averaging over angles yields $({\vec p}_K+{\vec p}_N)^2 \to (p_K^2+p_N^2)$. 
For bound hadrons, with $E_K=m_K-B_K,~E_N=m_N-B_N$, we expand near threshold, 
$E_{\rm th}=m_K+m_N$, neglecting quadratic terms in the binding energies 
$B_K, B_N$:  
\begin{equation} 
\sqrt{s}\approx E_{\rm th} - B_N - B_K - \frac{m_N}{m_N+m_K}\:
\frac{p_N^2}{2m_N} - \frac{m_K}{m_N+m_K}\:\frac{p_K^2}{2m_K}. 
\label{eq:sqrts} 
\end{equation}  
For $K^-$ capture at rest, we further neglect the atomic $B_K$ with respect to 
$B_N$ and replace the $K^-$ kinetic energy $p_K^2/(2m_K)$ in the local density 
approximation by $-{\rm Re}\:V_{K^-}(\rho)$ which dominates over the $K^-$ 
Coulomb potential within the range of densities of interest. The neutron 
kinetic energy $p_N^2/(2m_N)$ is approximated in the Fermi gas model by 
$23\,(\rho/\rho_0)^{2/3}$ MeV. Altogether the energy argument of the 
$K^- n \rightarrow \pi^- \Lambda$ BR assumes the form{\footnote{Applications 
of this form to kaonic atoms will be discussed elsewhere \cite{CFGGM11}.}} 
\begin{equation} 
\sqrt{s} \approx E_{\rm th} - B_N - 15.1\,(\rho/\rho_0)^{2/3} + 
0.345\,{\rm Re}\,V_{K^-}(\rho) \,\,\, {\rm (in~MeV)} 
\label{eq:BR(s)} 
\end{equation} 
which unambiguously prescribes the subthreshold two-body energy as a function 
of nuclear density at which BR($\sqrt{s},\rho$) of Fig.~\ref{fig1} is to 
be evaluated.{\footnote{Related ideas on the relevance of extrapolating to 
subthreshold energies in $K^-$ capture at rest have been repeatedly made by 
Wycech, see Ref.~\cite{wycech10}.}} Note that Eq.~(\ref{eq:BR(s)}) leads to 
implicit density dependence of BR($\sqrt{s},\rho$) through the invariant 
energy variable $\sqrt{s}$, in addition to the explicit $\rho$ dependence. 
The input BRs for our $1s_{\Lambda}$ hypernuclear formation rates calculation 
were obtained by averaging this chiral-model BR($\sqrt{s},\rho$), for a given 
$V_{K^-}(\rho)$, over the $K^-$ nuclear density overlap 
$\rho(r)\mid\Psi_{nLM}({\bf r})\mid^{2}$ of Eq.~(\ref{eq:rhobar}): 
\begin{equation} 
{\overline{\rm BR}} = \frac{1}{\overline \rho} \int {\rm BR}(\sqrt{s},\rho)\:
\rho(r) \mid \Psi_{nLM}({\bf r})\mid^{2} \:d^3r. 
\label{eq:BRbar} 
\end{equation} 
For $B_N$ we used target neutron separation energies. The nuclear 
densities used were obtained from modified harmonic oscillator 
nuclear charge densities by unfolding the finite size of the proton. 
The structure of Eqs.~(\ref{eq:BR(s)}), (\ref{eq:BRbar}), together 
with the plots of Fig.~\ref{fig1}, imply that deep $K^-$ nuclear 
potentials lead to significantly higher values of ${\overline{\rm BR}}$ 
than the threshold value used in Ref.~\cite{KCG10}, which indeed 
is borne out by the present calculations.

\section{Confronting data with calculations} 
\label{sec:DWIA} 


The $1s_{\Lambda}$ formation rates for a shallow $K^-$ nuclear potential 
$V_{K^-}^{\rm SH}$ of depth $-{\rm Re}\,V_{K^-}^{\rm SH}(\rho =\rho_0)\approx 
50$ MeV and for a deep $K^-$ nuclear potential $V_{K^-}^{\rm DD}$ of depth 
$-{\rm Re}\,V_{K^-}^{\rm DD}(\rho =\rho_0) \approx 190$ MeV have been 
recalculated with refined $K^-$ atomic wavefunctions and $\pi^-$ distorted 
waves.{\footnote{The complex $K^-$ nuclear potentials $V_{K^-}^{\rm SH}$ and 
$V_{K^-}^{\rm DD}$ were denoted $K_{\chi}$ and $K_{\rm DD}$, respectively, 
in Ref.~\cite{KCG10} where a complete listing of their parametrization is 
available.}} 
A major change here with respect to Refs.~\cite{KCG10,CFGK10v1} is the use 
of energy and density dependent BRs as outlined in Sect.~\ref{sec:BR}. 
The resulting ${\overline{\rm BR}}$s for the deep $K^-$ potential 
$V_{K^-}^{\rm DD}$ display considerable $A$ dependence, with values 
higher than the threshold value used in Ref.~\cite{KCG10}, particularly 
from ${^{12}{\rm C}}$ on. In contrast, the ${\overline{\rm BR}}$s for the 
shallow potential $V_{K^-}^{\rm SH}$ show little $A$ dependence, with values 
lower than the threshold value. Consequently, the difference between the DD 
and SH rates is no longer as large as calculated for a fixed BR threshold 
value \cite{KCG10}. For example, the calculated rates are (15--18)$\%$ of the 
experimentally derived rate for $^7$Li under procedure (a) using the best-fit 
pion optical potential $\pi_e$, Eqs.~(19), (20) of Ref.~\cite{CFGM01}, and 
(23--26)$\%$ of it using the pion optical potential $\pi_b$ employed in 
Ref.~\cite{KCG10}. 

\begin{figure}[htb] 
\begin{center} 
\includegraphics[width=0.7\textwidth]{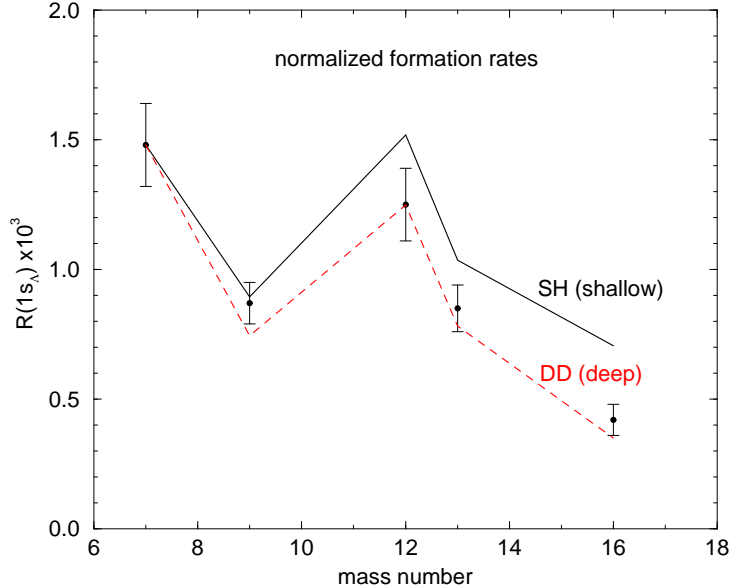} 
\caption{Comparison between $1s_{\Lambda}$ formation rates derived from 
the FINUDA $K^-$ capture at rest data \cite{agnello10,agnello05} and DWIA 
calculations normalized to the $1s_{\Lambda}$ formation rate of 
$_{\Lambda}^{7}{\rm Li}$ listed in Table~\ref{tab1} for shallow (SH, solid) 
and deep (DD, dashed) $K^-$ nuclear potentials. The calculated $1s_{\Lambda}$ 
formation rates use $K^- n \to \pi^- \Lambda$ in-medium BRs without self 
energies, see Sect.~\ref{sec:BR}, and pion optical potential $\pi_e$ from 
Ref.~\cite{CFGM01}. The error bars consist of statistical uncertainties only.} 
\label{fig2} 
\end{center} 
\end{figure} 

Here we focus on the $A$ dependence of the $1s_{\Lambda}$ formation rates. 
For given $K^-$ and $\pi^-$ potentials the calculated rates are scaled up 
by a normalization factor to achieve agreement for ${^7{\rm Li}}$ 
with the $1s_{\Lambda}$ rate derived from the data under procedure (a) in 
Table~\ref{tab1}. This is shown in Fig.~\ref{fig2} where the uncertainties 
of the experimentally derived $1s_{\Lambda}$ rates consist only of statistical 
errors that vary from one target to another. The systematic errors, on the 
other hand, are the same for all targets and drop out when considering $A$ 
dependence within the present set of FINUDA data. The normalized calculated 
$1s_{\Lambda}$ rates shown in the figure are based on ${\overline{\rm BR}}$s 
calculated according to Eq.~(\ref{eq:BRbar}) from the BRs plotted on the 
l.h.s. of Fig.~\ref{fig1} (CS30, no-SE). Results are shown for the pion 
optical potential $\pi_e$ which was fitted to $\pi^{-}$--${^{12}{\rm C}}$ 
angular distributions at 162 MeV \cite{CFGM01}, and  for the two $K^-$ 
nuclear potentials $V_{K^-}^{\rm SH}$ and $V_{K^-}^{\rm DD}$. We note that 
the decrease of the experimentally derived $1s_{\Lambda}$ rates from 
${^7{\rm Li}}$ to ${^9{\rm Be}}$, followed by increase for ${^{12}{\rm C}}$ 
and subsequently decreasing through ${^{13}{\rm C}}$ down to ${^{16}{\rm O}}$, 
is well reproduced by both calculations shown in Fig.~\ref{fig2}. However, the 
deep $V_{K^-}^{\rm DD}$ calculated rates reproduce better the $A$ dependence 
of the experimentally derived rates than the shallow $V_{K^-}^{\rm SH}$ 
potential does. In reaching this conclusion on $V_{K^-}^{\rm DD}$, the 
increase of the $K^-n\rightarrow\pi^-\Lambda$ $\overline{\rm BR}$ values 
between ${^7{\rm Li}}$ and ${^{16}{\rm O}}$ is essential, by moderating the 
fall off of the rates calculated using $A$ independent $\overline{\rm BR}$s. 
Similar conclusions hold for the $A$ dependence of rates calculated using 
${\overline{\rm BR}}$s that are derived according to Eq.~(\ref{eq:BRbar}) 
from the BRs plotted on the r.h.s. of Fig.~\ref{fig1} (CS30, with SE). 
On the other hand, if the pion optical potentials $\pi_b$ or $\pi_c$ 
(applied in Ref.~\cite{KCG10}) are used in these calculations, neither 
$V_{K^-}^{\rm DD}$ nor $V_{K^-}^{\rm SH}$ do as good a job as the combination 
$V_{K^-}^{\rm DD}$ and the best-fit $\pi_e$ does, and no firm conclusion can 
be drawn.

\section{Conclusion} 
\label{concl} 

In conclusion, we have derived $1s_{\Lambda}$ hypernuclear 
formation rates from peak formation rates associated with the 
$^{A}{\rm Z}(K^-_{\rm stop},\pi^-){{^{A}_{\Lambda}}\rm{Z}}$ spectra 
presented recently by the FINUDA Collaboration on several nuclear targets 
in the $p$~shell \cite{agnello10,agnello05}. We then compared the $A$ 
dependence of these derived rates with that provided by calculations 
for the two extreme $V_{K^-}$ scenarios discussed at present, 
a shallow potential \cite{RO00,CFGM01} and a density dependent deep 
potential \cite{FG07}. The calculations use $K^- n \to \pi^- \Lambda$ 
in-medium BRs generated by applying a recent chirally motivated coupled 
channel model \cite{CS10}. These BRs exhibit a strong subthreshold energy 
and density dependence, as shown in Fig.~\ref{fig1}, and therefore result 
in $A$ dependent input values $\overline{\rm BR}$ that depend sensitively 
on the initial-state $K^-$ nuclear potential $V_{K^-}$. The calculations 
also demonstrate additional strong sensitivity to $V_{K^-}$ through 
the atomic wavefunctions it generates which enter the DWIA amplitude 
Eq.~(\ref{eq:DW}), as discussed extensively in previous calculations 
of $K^-_{\rm stop}$ hypernuclear formation rates \cite{CFGM01,KCG10}. 
The comparison between the calculated $A$ dependence and that derived 
from the FINUDA data slightly favors a deep $K^-$ nuclear potential over 
a shallow one. This conclusion outdates the one reached in an earlier version 
in which the energy and density dependence of the BRs, resulting here in 
a new source of sensitivity to $V_{K^-}$, was disregarded \cite{CFGK10v1}. 
In future work, it would be interesting to use other versions of $K^-N$ 
chirally motivated models and to extend the range of nuclear targets used in 
stopped $K^-$ reactions to medium and heavy weight nuclei in order to confirm 
the present conclusion, and to look for more subtle effects of density 
dependence.

\section*{Acknowledgements} 
We are grateful to Tullio Bressani and Germano Bonomi for useful discussions 
of the FINUDA measurements and to Daniel Gazda and Ji\v{r}\'{i} Mare\v{s} for 
clarifying discussions on subthreshold extrapolations. This work was supported 
by the GAUK Grant No. 91509 and GACR Grant No. 202/09/1441, as well as by the 
EU initiative FP7, HadronPhysics2, under Project No. 227431.

\end{document}